\documentclass[11pt,twoside]{article}

\usepackage{asp2006}

\begin{document}
\title{Potential Targets for ASTRO-G In-Beam Phase-Referencing} 
\author{S. Frey,$^{1,2}$ and K.\'E. Gab\'anyi$^{2,1,3}$}
\affil{
$^1$ F\"OMI Satellite Geodetic Observatory, P.O. Box 585, H-1592 Budapest, Hungary\\
$^2$ MTA Research Group for Physical Geodesy and Geodynamics, P.O. Box 91, H-1521 Budapest, Hungary\\
$^3$ Institute of Space and Astronautical Science/JAXA, 3-1-1 Yoshinodai, Sagamihara, Kanagawa 229-8510, Japan\\
}

\begin{abstract} 
We show that as many as $\sim$50 quasars with at least mJy-level expected flux density can be pre-selected as potential in-beam phase-reference targets for ASTRO-G. Most of them have never been imaged with VLBI. These sources are located around strong, compact calibrator sources that have correlated flux density $>100$~mJy on the longest VLBA baselines at 8.4~GHz. All the targets lie within $12\arcmin$ from the respective reference source. The basis of this selection is an efficient method to identify potential weak VLBI target quasars simply using optical and low-resolution radio catalogue data. The sample of these dominantly weak sources offers a good opportunity for a statistical study of their radio structure with unprecedented angular resolution at 8.4~GHz.

\end{abstract}

\section{Introduction}

Phase-referencing is a way to increase the sensitivity of the Space Very Long Baseline Interferometry (SVLBI) observations that provide extremely high angular resolution due to the baselines exceeding the Earth diameter. Phase-reference imaging in ground-based VLBI is usually done in cycles of interleaving observations between a weak target source and a nearby strong reference source. Delay, delay-rate and phase solutions obtained for the phase-reference calibrator are interpolated and applied for the target source within the atmospheric coherence time, thus increasing the coherent integration time on the weak target source.   

Unlike the first dedicated SVLBI satellite HALCA \citep{hira00}, the next-generation satellite ASTRO-G will be capable of rapid attitude changing maneuvers. This, and the accurate orbit determination will allow us to observe suitable nearby reference--target source pairs in the traditional ``nodding'' style \citep{asak07}. There is another, technically less demanding method which does not require rapid changes in the space antenna pointing if the reference--target separation is so small that both sources are within the primary beam of the 9.3-m ASTRO-G paraboloid antenna ($\sim12\arcmin$ at 8.4~GHz). In this scenario, the ground-based part of the SVLBI network performs the usual reference--target switching cycles, while the space antenna remains pointed to the same celestial position. (The diameters of the ground-based VLBI antennas are at least a factor of $\sim3$ larger, and thus their primary beam sizes are considerably smaller than that of the orbiting antenna.)

Successful in-beam phase-referencing experiments have already been conducted with HALCA which could not quickly change its antenna pointing \citep{pori00,bart00,porc00,guir01}. However, the use of in-beam phase-referencing is severely limited by the small number of sufficiently close source pairs known in the sky. Generally speaking, for any given target source of interest, it is very unlikely to find a suitable phase-reference calibrator within the primary beam of even the relatively small-diameter space antenna.

One may reverse the usual logic and select the phase-reference calibrator sources first. {\it Then} it becomes possible to look for potential weak target sources that are located so close to one of the reference sources that in-beam phase-referencing observations with the orbiting antenna are feasible. Here we show that as many as $\sim$50 quasars with at least mJy-level expected flux density can be pre-selected as potential in-beam phase-reference targets for ASTRO-G. This prospective sample is large enough for a statistical study of sub-milliarcsecond radio structures of weaker sources in comparison with bright ones. Such a sample, most of which have never been studied with VLBI, would certainly contain individually interesting radio quasars as well. The suitability of at least some of the candidate sources could be verified with ground-based VLBI observations prior to the launch of ASTRO-G.

\section{Sample selection}

Potential phase-reference sources are easily found in the most complete and up-to-date NASA GSFC VLBI source catalogue. The version we used here is 2007a\_astro \citep{petr07} which incorporates the VLBA Calibrator Survey (VCS) sources \citep{beas02,foma03,petr05,petr06,kova07}. To pick up the compact and bright objects that are most likely to give good signal-to-noise ratio with SVLBI, we applied a lower limit of 100~mJy for the 8.4-GHz correlated flux density at the longest ground baselines. Nearly 1900 sources passed this filtering.

The next step was to look for other objects within $12\arcmin$ of the positions of each potential reference source. This angular separation equals to the primary beam size (HPBW) of the ASTRO-G antenna at 8.4~GHz. In other words, the reference sources and the targets -- if found at all -- lie within this beam. The search was performed in the Sloan Digital Sky Survey (SDSS) Data Release 6 (DR6) data base\footnote{\tt http://www.sdss.org/dr6} \citep{adel08}. The choice of an optical catalogue may seem odd at the first glance, but our experience with the Deep Extragalactic VLBI-Optical Survey \citep[DEVOS,][]{moso06,frey08} shows that 85\% of the SDSS optical {\it quasars} (i.e. extragalaxtic objects with stellar appearance) that coincide with an unresolved ($<5\arcsec$) and ``strong'' ($>20$~mJy) radio source in the VLA Faint Images of the Radio Sky at Twenty-centimeters (FIRST) Survey list\footnote{\tt http://sundog.stsci.edu} \citep{whit97} are detected with phase-referenced ground-based VLBI at 5~GHz. Therefore the cross-comparison of the SDSS and FIRST lists provides us with an efficient tool to pick up potential VLBI target sources at cm wavelengths, with at least mJy-level correlated flux densities. These sources have optical magnitudes readily available from SDSS.

\section{Results}

We found a total of 62 objects (quasars) which are unresolved in {\it both} optical and with the VLA in FIRST. Only a few of them are known as relatively bright close pairs of VLBI sources (e.g. J1300+141A and J1300+141B). The majority of the objects found have never been studied with mas-resolution radio imaging. Considering the DEVOS detection rate at 5~GHz \citep{frey08}, we estimate that a fairly large sample, about 50 such quasar--reference source pairs could be successfully targeted with in-beam phase-referenced SVLBI observations at 8.4~GHz. Note that many more similar pairs should exist in the sky. However, the SDSS and FIRST catalogues have a limited sky coverage therefore our method cannot be applied to identify them.

\acknowledgements
This research was partly supported by the Hungarian Scientific Research Fund (OTKA T046097) and the Hungarian Space Office (TP-314). K\'EG acknowledges a fellowship received from the Japan Society for Promotion of Science.


\end{document}